\documentclass[runningheads]{svmult}

\usepackage{makeidx}   
\usepackage{graphicx}  
\usepackage{subeqnar}  
\usepackage{multicol}  
\usepackage{physprbb}  
\makeindex             

\def\arcsec{\hbox{$^{\prime\prime}$}}

\def\ha{H$\alpha$}

\font\sc=cmr7

\def\HII   {H{\sc II}}

\begin{document}
\title*{Spectroscopy of faint emission line nebulae}
\toctitle{Spectroscopy of faint emission line nebulae}
%
%
\titlerunning{Spectroscopy of faint emission line nebulae}
%
\author{Ralf-J\"urgen Dettmar}

\authorrunning{R.-J. Dettmar}
%
%
\institute{Astronomical Institute, Ruhr-University Bochum, 44780 Bochum, Germany}

\maketitle              

\begin{abstract}

Based on results obtained with FORS1 in long-slit mode we demonstrate the
power of the VLT for the spectroscopy of faint emission line nebulae such as
the gaseous halos of diffuse ionized gas (DIG) in spiral galaxies. It is
shown that VLT spectra of DIG allow us to address the ionization and
excitation processes for the interstellar medium on galactic scales and 
that in the future more detailed kinematical studies of DIG  could help 
to constrain the origin of the observed thick 
H$^+$ layer in galaxies. 

The currently available instrumentation with regard to this application is
compared to other possible designs and efficiencies for integral field
spectrographs, in particular Fabry-Perot systems.
\end{abstract}

\section{Introduction}
Emission lines originating in ionized gas provide very valuable diagnostics
for  physical conditions in several components of the interstellar medium 
(ISM) such as \HII \ regions, supernova remnants (SNRs), planetary nebulae, 
or the warm ionized medium. To demonstrate the complex relation of these 
various components we show in Fig. 1 the \ha \ emission of the central
star-forming region of the nearby late-type galaxy NGC\,55. 
Besides the presence of localized sources (i.e., \HII \ regions and SNRs) 
this image demonstrates that a significant fraction of \ha \ emission 
originates outside of ``classical" \HII-regions (which are saturated black 
in the grey scale representation) in shells, loops, bubbles, filaments, and 
knots, as well as a smoothly distributed, diffuse component. This  H$^+$ gas 
outside of \HII \ regions is now frequently called (despite its 
morphological diversity) {\it Diffuse Ionized Gas} (DIG) or {\it Warm 
Ionized Medium} (WIM) and  can be identified with the {\it Reynolds-layer} 
of the Milky Way. The \ha \ emission of the Galaxy  is currently being mapped 
in a northern sky survey by the Wisconsin H-Alpha Mapper (WHAM; Reynolds et 
al. 1998) and first results from WHAM (Haffner et al. 1998) show striking 
similarities between the DIG in NGC\,55 (Fig. 1) and the Galaxy. 
A more complete documentation demonstrating the power of the WHAM 
Fabry-Perot survey for Galactic studies can be found at [9].

\begin{figure}[]
\begin{center}
\includegraphics[width=.6\textwidth,angle=-90]{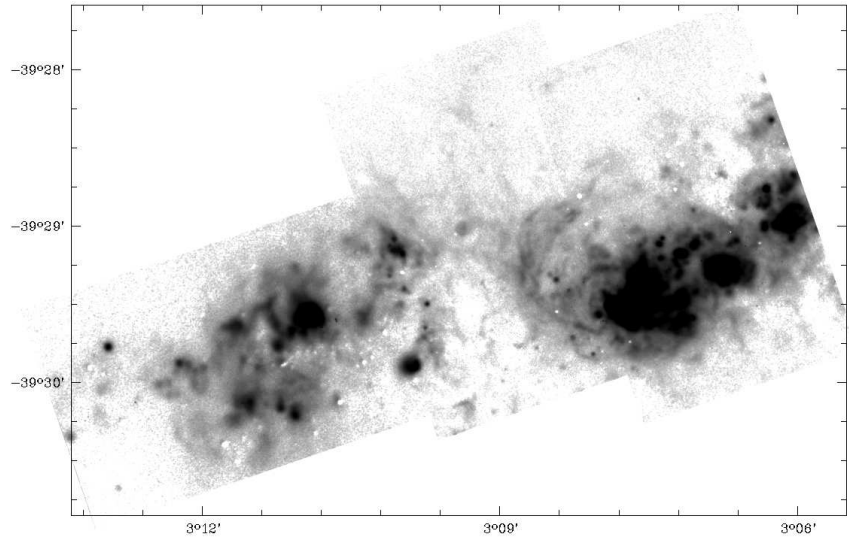}
\end{center}
\caption[]{\ha \ image of the central part of NGC\,55 adopted from 
Otte and Dettmar (1999).}
\label{fig:n55}
\end{figure}

The filamentary structure  of \ha \ emitting DIG in Fig. 1 can be traced
out into the halo on scales of several hundreds of pc and becomes an 
observational challenge with increasing distance from the mid-plane and 
decreasing flux. This faint extra-planar DIG corroborates scenarios of a 
large scale exchange of matter between the ISM in galactic disks 
and their halos driven by the energy input of star forming regions in the 
disk. More detailed reviews of this aspect can be found in, e. g., Dahlem 
(1997) or Dettmar (1999).

\section {Detection of very extended DIG halos of spiral galaxies}

The first imaging detection of a $\sim$1\,kpc thick layer of DIG in NGC\,891 
(Rand et al. 1990, Dettmar 1990)
was meanwhile confirmed spectroscopically (Rand 1997) extending the 
detection limit out to at least 5\,kpc  above the plane.
The most active galactic disks with regard to their star formation rate
 possess spectacular gaseous halos, e.g, in the case
of NGC\,4631 an extend of 16\,kpc in radius has been claimed from a narrow band
imaging experiment  by Donahue et al. (1995).
Using FORS1 in long-slit mode we could recently establish {\it by spectroscopy}
(T\"ullmann et al. 2000) that also the DIG halo in NGC\,5775 extents out to  at least 10\,kpc (Fig. 2),
confirming and extending the work by Rand (2000).

This finding is in itself of interest in the context of the chemical evolution
history of disk galaxies, since the gaseous halo and in particular its
hot component could transport and maintain significant amounts of metals.
The detailed physics of such  mass exchanges or
outflows are of course most important for the  understanding of the metal 
enrichment processes in early phases of galaxy evolution. It is therefore
of some importance that not even the energy balance of the ISM is well
understood: the ionization and excitation conditions of the DIG component 
require more than photoionization by OB stars can directly supply. However,
it is unclear what additional process(es) 
(turbulence, dust heating, shocks, magnetic
reconnection) contribute(s) to the heating, in particular since there is 
more and more observational evidence that the temperature is increasing
with height above the plane {\it z} (Reynolds et al. 1999, 
for the Milky Way; Rand 1997, for NGC\, 891; T\"ullmann and
Dettmar 2000, for a sample based on ESO/La Silla observations). 
Further progress with regard to the physical conditions of the ionized
halo gas would require deep spectra that allow for the detection of faint
diagnostic emission lines over a large wavelength range. 
This problem can be best treated with a specialized
high-efficiency spectrograph and is addressed by Prieto (2001) elsewhere
in this volume.

\begin{figure}[]
\begin{center}
\includegraphics[width=.6\textwidth]{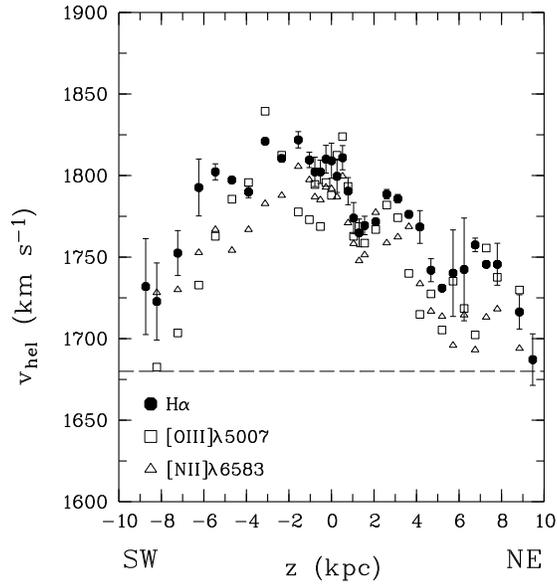}
\end{center}
\caption[]{Heliocentric velocities perpendicular to the disk (at $\sim30$\arcsec
 \ SE of the nucleus) of NGC\,5775 
for prominent emission lines extracted from a VLT FORS1 spectrum. The noise level of $1.26\times\,10^{-19}$\,ergs cm$^{-2}$\,s$^{-1}$\,\AA$^{-1}$ reached in 3 hrs allows for a detection out
to 10\,kpc above the disk. The dashed line indicates the systemic velocity (from T\"ullmann et al. 2000).}
\label{fig:rot}
\end{figure}

\section{Kinematics of the extraplanar DIG in NGC\,5775}

The deep VLT spectra of NGC\,5775 are, however, of interest also with respect
to the kinematical information as 
Fig.\,2 shows that  the rotational velocity of the DIG halo drops
to systemic high above the plane. In addition, it was shown (T\"ullmann et al. 2000), that  
the DIG halo is associated with highly ordered magnetic fields, surprising
in view of the energy input (i.e., turbulence, flows etc.) from the
star formation activity in the underlying disk. 

These findings now allow us to discuss some new physical processes
to explain the ionization/excitation as well as the kinematics of the 
halo gas.
The surprising kinematical information can be used to study the 
(magneto?)hydro$\-$dynamics
of a possible large scale outflow. Since the observed drop in velocity is
dramatic, one has to expect that the shape of the dark matter distribution
plays an important role, too. Since the magnetic field structure with a strong
vertical component also is very suggestive for outflows -- such a magnetic
field structure actually would favour outflow rather than suppressing it --
and the question of the gas metalicities in the halo becomes even more
important for chemical enrichment processes.  

Another interesting aspect is added by the recent finding from an UV 
absorption line study of galactic halos. C\^ot\'e et al. (2000) report 
that absorption lines
associated with galactic halos are observed at the systemic velocity of the
host galaxy independent of the impact parameter for the line-of-sight.   

\begin{figure}[h]
\begin{center}
\includegraphics[width=.6\textwidth,angle=-90]{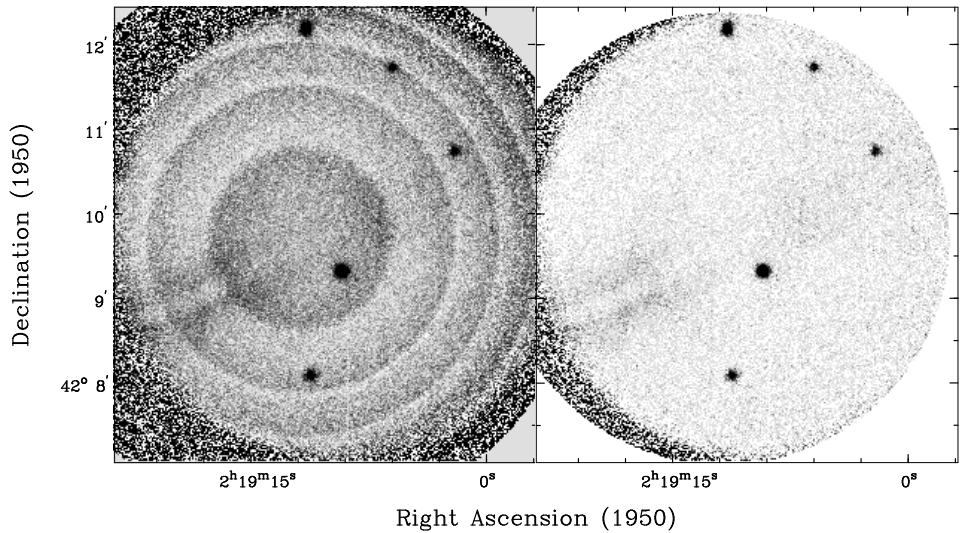}
\end{center}
\caption[]{Wavelength resampled channel of a TAURUS data cube of NGC\,891
before (left) and after (right) removal of interfering night sky line (from
Dettmar, Allen, \& van der Hulst).}
\label{fig:FP}
\end{figure}

\section{The need for high-resolution integral field spectroscopy}

For both applications mentioned in the two previous sections 
-- the detection of faint emission and the extraction of kinematical 
information --  a higher
spectral resolution than typically reachable with grism spectrographs
would be an advantage. Considering the manifold of integral field spectrographs already available at the VLT and discussed at this meeting as possible 
future VLT instruments, a Fabry-Perot design may best meet this requirements, 
in particular if good spatial coverage is another constraint. 
The power of a Fabry-Perot spectrograph is well
demonstrated in several applications by Bland-Hawthorn (see e.g. 
Bland-Hawthorn et al. 1997) and the more specialist use of a Fabry-Perot as a tunable filter
 attached
to the VLT was recently discussed as a modification to FORS (Jones et al. 2001). 

To demonstrate the gain in S/N for the detection of emission lines in the
presence of bright night sky lines by using a Fabry-Perot 
we reproduce in Fig. 3 a channel 
of a TAURUS data cube of NGC\,891 before and after correction for
the night sky contribution. With an  integration time of 2\,min per channel
 a detection of $\sim$ 2$\times$10$^{-17}$\,ergs 
cm$^{-2}$\,s$^{-1}$\,arcsec$^{-2}$ could be reached at the 4m WHT.   

\section{Conclusions}

The examples of NGC\,5775 and NGC\,4631 given above demonstrate that the 
extent of ionized gaseous halos around spiral galaxies is limited by the
detection limit reached. Also kinematical studies would much benefit 
from higher
spectral resolution over a large field of view, with  R between 10000 and 
30000, depending on the
scientific case. A scanning Fabry-Perot spectrograph seems to be a very
versatile instrument to meet these requirements and compared favourable 
to other integral field spectrometer designs
with regard to spacial sampling and spectral resolution. It would therefore
ideally complement the current VLT instrumentation.\\

\noindent{\it Acknowledgements.} Its a pleasure to thank Ron Allen and
Thijs van der Hulst for the patient collaboration on our ``long term'' 
Fabry-Perot  project. 
Thanks also to staff and students at Ruhr-University for contributing
the ``real'' work presented here. The author acknowledges partial 
financial support  in this field by DFG  through SFB\,191.

%

\end{document}